# Tunnel spectroscopy of localised electronic states in hexagonal boron nitride

M.T. Greenaway [1,2], E.E. Vdovin[1,3], D. Ghazaryan[4], A. Misra[4], A. Mishchenko[4], Y. Cao[5], Z. Wang[4], J.R. Wallbank[5], M. Holwill [5], Yu.N. Khanin[3], S.V. Morozov[3,6], K. Watanabe [7], T. Taniguchi[7], O. Makarovsky[1], T.M. Fromhold[1], A. Patanè[1], A.K. Geim[4,5], V.I. Fal'ko [5], K.S. Novoselov[4,5] & L. Eaves[1,4]

Hexagonal boron nitride is a large band gap layered crystal, frequently incorporated in van der Waals heterostructures as an insulating or tunnel barrier. Localised states with energies within its band gap can emit visible light, relevant to applications in nanophotonics and quantum information processing. However, they also give rise to conducting channels, which can induce electrical breakdown when a large voltage is applied. Here we use gated tunnel transistors to study resonant electron tunnelling through the localised states in few atomic-layer boron nitride barriers sandwiched between two monolayer graphene electrodes. The measurements are used to determine the energy, linewidth, tunnelling transmission probability, and depth within the barrier of more than 50 distinct localised states. A three-step process of electron percolation through two spatially separated localised states is also investigated.

[1] School of Physics and Astronomy, University of Nottingham, Nottingham NG7 2RD, UK. [2] Department of Physics, Loughborough University, Loughborough LE11 3TU, UK. [3] Institute of Microelectronics Technology and High Purity Materials, RAS, Chernogolovka 142432, Russia. [4] School of Physics and Astronomy, University of Manchester, Manchester M13 9PL, UK. [5] National Graphene Institute, University of Manchester, Manchester M13 9PL, UK. [6] National University of Science and Technology "MISiS", 119049Leninsky pr. 4, Moscow, Russia. [7] National Institute for Materials Science, Namiki 1-1, Tsukuba, Ibaraki 305-0044, Japan. Correspondence and requests for materials should be addressed to M.T.G. (email: m.t.greenaway@lboro.ac.uk) or to L.E. (email: laurence.eaves@nottingham.ac.uk)

Following the discovery of the remarkable electronic properties of graphene[1], researchers have investigated a variety of other layered crystalline compounds that remain chemically stable even when they are disassembled into atomically thin flakes by mechanical exfoliation[2–4]. These two-dimensional crystals can be stacked to form heterostructures and functional devices[5–18] held together by van der Waals (vdW) forces that preserve the structural integrity and physical properties of the component layers. Of particular significance is hexagonal boron nitride (hBN)[2,19], which has a lattice constant only 1.8% larger than that of graphene. It is a large band gap material that can be used as an insulating barrier for a gating electrode, as a barrier for tunnelling electrons, or as a source of ultraviolet light[2].

Recent studies have demonstrated that crystals of hBN can contain strongly localised electronic states within the energy gap[20–27]. These states are attributed to the presence of structural defects and impurities likely to be present even in nominally pure hBN crystals. They could also be introduced unintentionally during mechanical exfoliation of hBN and/or its incorporation within a multilayer vdW heterostructure. Electronic transitions between localised states with energies within the large band gap of hBN are also of interest, because they are single quantum emitters of visible light[28–40] and thus have potential for applications in nanophotonics, optoelectronics and quantum information processing. Recently, localised states have been shown to affect the electronic properties of spintronic[41] and superconducting[42] van der Waals devices. Defect-related phenomena can also impair the electrical properties of future devices based on hBN by inducing random telegraph noise and causing electrical breakdown of its insulating properties when a sufficiently strong electric field is applied[43,44].

In this paper, we investigate how electrons tunnel resonantly between two monolayer graphene electrodes through localised states within an hBN barrier. Our devices incorporate either one or two gate electrodes, which provide precise control of the density and chemical potentials of the carriers in the graphene layers. The measurements allow us to determine the energy and spatial position of each of the localised states. The crystalline lattices of the two monolayer graphene electrodes are misaligned by a small angle of a few degrees. This twist angle suppresses direct band-to-band resonant tunnelling where the in-plane momentum component of the tunnelling electron is conserved[8,10–12], and helps resolve clearly the small tunnel current passing through an individual localised state. The momentum conservation rule is relaxed for the case of tunnelling through the bound states within the band gap of hBN due to their strong spatial localisation[17,18].

## Results

**Resonant tunnelling through a single localised state**. The top left inset of Fig. 1a is a schematic diagram showing the configuration of Device 1. A few atomic layers of hBN (green) forms a tunnel barrier sandwiched between two graphene monolayers, $Gr_b$ and $Gr_t$, which act as source and drain electrodes. The application of a bias voltage, $V_b$, between them causes a tunnel current, $I$, to flow through the hBN barrier. A third graphite layer ($Gr_g$), which lies on a $SiO_2$ substrate, is separated from $Gr_b$ by an insulating hBN layer. This gate electrode is used to adjust the carrier sheet density of the graphene layers by varying the gate voltage, $V_g$. The active area for current flow in Device 1 is ~50 μm². Further details of the device fabrication are given in the Methods section and ref. [10].

The red curve in Fig. 1a shows the $I(V_b)$ curve at a measurement temperature, $T = 1.75$ K, and $V_g = 0$. For $|V_b| \lesssim 200$ mV, the tunnel current is small, but has a step-like increase when $V_b = V_1 \approx \pm 200$ mV. The differential conductance plot, $G = dI/dV_b$, shown in Fig. 1b, displays the increase of current at the step edge as a sharp peak. At $V_g = 0$, the two strong and sharp peaks at $V_b \approx \pm 200$ mV are accompanied by weaker features at higher $|V_b|$. We attribute each of the two strong peaks to the threshold of resonant tunnelling through the same localised state (state A) within the hBN barrier when its energy, $E_A$, becomes aligned with one or other of the chemical potentials, $\mu_b$ or $\mu_t$, of the bottom (b) or top (t) graphene layers. For $V_b > V_1$, the conductance channel through the localised state remains open, see lower right inset of Fig. 1a for this general case. An increase of $V_g$ decreases the $|V_b|$ position of the two conductance peaks until at $V_g = 1.7$ V they merge into a single peak centred at $V_b = 0$, see the green and blue curves in Fig. 1a, b.

Figure 2a is a colour map of $G(V_b, V_g)$ measured for Device 1. The white curves show a series of seven $G(V_b)$ plots at selected $V_g$ (5 V, 3 V, ..., −7 V). Close to the top of Fig. 2a, the white arrows highlight the positions of the two strong peaks in $G(V_b)$ at $V_g = 5$ V. The $V_b$ positions of the peaks are strongly dependent on $V_g$: over the range of $V_g$ from +7 V through 0 to −7 V, their loci have a prominent X-shaped dependence, corresponding to the onset of electron tunnelling through state A.

We reproduce these measurements accurately using the Landauer-Büttiker conductance formula[45–48], combined with Fermi's golden rule and an electrostatic model of the device, Fig. 2b. It includes the quantum capacitance of graphene that arises from its low density of states near the Dirac points. Details

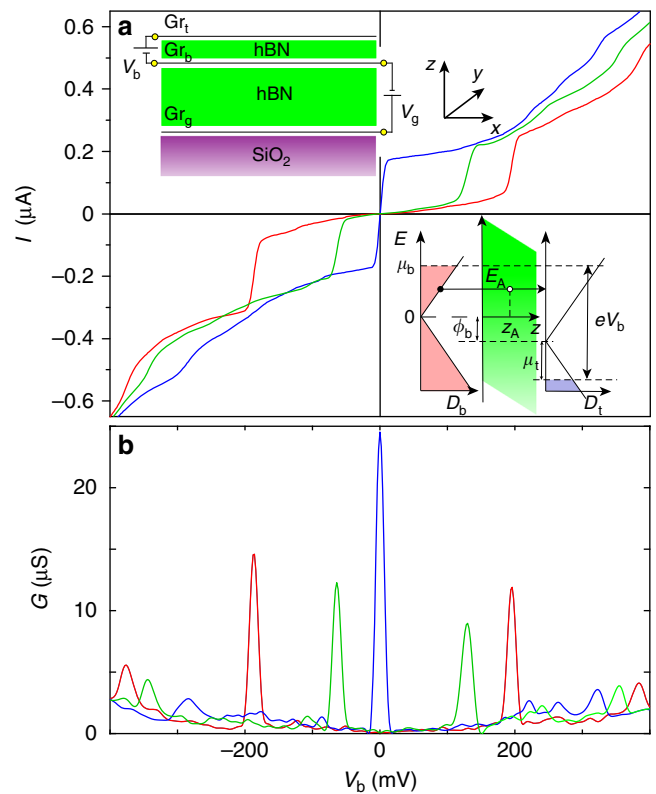

**Fig. 1** Current-bias voltage characteristics of Device 1. **a** Measured current-bias voltage curves, $I(V_b)$, when the gate voltage, $V_g = 0$ (red), $V_g = 0.8$ V (green) and $V_g = 1.7$ V (blue) ($T = 1.75$ K). Top left inset: schematic diagram of the device showing the graphite gate ($Gr_g$) and electrodes through which the current flows ($Gr_{b,t}$) (horizontal black lines), hBN layers (lime green), the voltage configuration and definition of spatial axes. Bottom right inset: band diagram showing the densities of states $D_b$ and $D_t$ in the bottom and top graphene layers and their chemical potentials, $\mu_b$ and $\mu_t$. The electron tunnels through a localised state 'A' of energy $E_A$ and spatial position $z_A$, see open circle. **b** Differential conductance, $G = dI/dV_b$ of the $I(V_b)$ curves in **a**

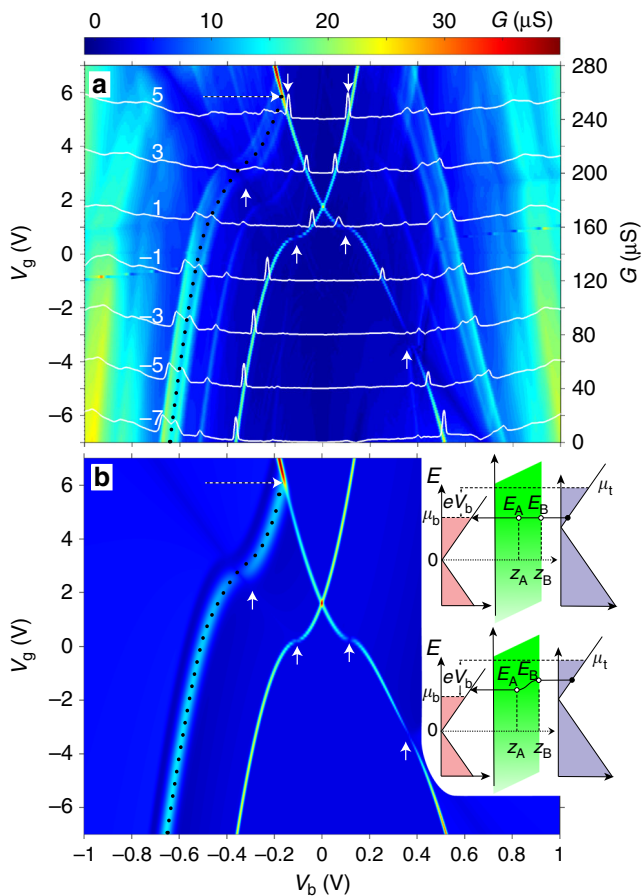

**Fig. 2** Gate and bias voltage dependence of differential conductance for Device 1. **a** Colour map showing the dependence of the differential conductance, $G = dI/dV_b$ on bias, $V_b$, and gate, $V_g$, voltages, white curves show $G(V_b)$ when $V_g = -7$ to 5 V in 2 V intervals, scale right, with curves offset by 40 μS. **b** A fit to the measurements using the theoretical model described in text and Supplementary Notes 1 and 2. Upward arrows highlight gaps in the loci arising from alignment of the chemical potential in the two graphene layers with the low density of states at the Dirac point. Horizontal arrows highlight the position where the sequential tunnelling feature vanishes. The black dotted curves indicate the loci where states A and B are in energetic alignment, $E_B = E_A$, illustrated by the upper inset in **b**, the lower inset in **b** shows schematically the sequential inelastic tunnelling process when $E_B > E_A$

of the model are presented in Supplementary Notes 1 and 2, and refs. [7,9–11,49]. The measured device characteristics are described accurately by the model. In this way, we obtain accurate values for the tunnel barrier thickness, $d = 1.5$ nm, and lower insulating hBN layer thickness, $d_g = 25$ nm. The electric field, $F_b$, across the hBN tunnel barrier generated by the bias voltage-induced charge on the graphene layers shifts the energy level of a given localised state, A, so that its energy relative to the Dirac point of the bottom graphene layer is $E_A = E_A^0 + eF_b z_A$. To obtain a best fit to the data, we set the energy of the level at the flat band condition ($F_b = 0$) to be $E_A^0 = 0.11$ eV; we set its location to be in the middle of the barrier so that its spatial coordinate perpendicular to the plane of the layers and relative to the position of the bottom graphene layer is $z_A = d/2 = 0.75$ nm, see lower right inset of Fig. 1a.

At low bias ($V_b < V_1$) and low temperatures there are few electrons with sufficient energy to tunnel with energy conservation through the localised state. The bias voltage $V_b$ is given by $eV_b = \mu_b - \mu_t - \phi_b$, where $\phi_b = eF_b d$ and $\mu_b$ and $\mu_t$ are measured with respect to the Dirac points of the graphene electrodes. When $V_b$ is increased, $\mu_b$ increases and $E_A$ decreases. When $V_b = V_1$, $E_A = \mu_b$ so that electrons can tunnel with energy conservation through this localised state, thus opening a conduction channel between the two graphene layers, and producing the peak in $G$. Similarly, for $V_b < 0$, tunnelling through the same impurity can be achieved when $E_A$ aligns with the chemical potential in the top layer, i.e. $E_A = \mu_t + \phi_b$. The model provides an accurate fit to the measured data as can be seen by comparing our modelled conductance Fig. 2b with the measured data in Fig. 2a. Note the positions on the X-shaped loci at which the measured amplitude of the conductance peaks is suppressed; these are indicated by vertical white arrows in both maps. The model calculation in Fig. 2b confirms that this suppression occurs when the chemical potential in one or the other graphene electrode passes through its Dirac point where the density of states approaches zero. The good agreement between the measured and modelled zero conductance loci validates our electrostatic model.

We find that the peaks in $G$ broaden as $T$ increases, consistent with the thermal broadening of the electron energy distribution at the chemical potentials of the two graphene layers, see Supplementary Fig. 1. By comparing our model with the data over the temperature range from 1.75 to 90 K, we estimate the full width half maximum linewidth of the state to be $\gamma \approx 6$ meV and the lifetime $\hbar/\gamma \approx 0.1$ ps. The best fit to the data is obtained when we use a Gaussian lineshape, see Supplementary Note 2 for more details. This is consistent with studies of the lineshape of optical emission from localised states in hBN[37] and corresponds to inhomogeneous broadening[50] of the state. This lineshape could arise from spectral diffusion due to local electrostatic fluctuations in the vicinity of the state. A similar effect has also been reported for colour centres in diamond[51,52].

The peak in conductance at $V_b = 0$ when $V_g = 1.7$ V and $T = 1.75$ K corresponds to $G_p = \beta e^2/h$, where $e^2/h$ is the quantum of conductance and the measured parameter $\beta = 0.75$. For coherent tunnelling through a localised state with a Gaussian density of states $\beta = \sqrt{\pi \ln 2} S$, where $S = 4\gamma_b \gamma_t/(\gamma_b + \gamma_t)^2 \approx 0.5$ is the total transmission probability. Here, $\gamma_b/\hbar$ and $\gamma_t/\hbar$ are the electron tunnelling rates between the localised state and the b and t electrodes and $\gamma = \gamma_b + \gamma_t$[48]. Note that if $\gamma_b = \gamma_t$ then $S = 1$; however, in contrast we find that $\gamma_b \sim 0.8\gamma$ and $\gamma_t \sim 0.2\gamma$, which means that the state is somewhat more strongly coupled to the bottom layer than the top, see Supplementary Note 2 for more details.

**Sequential tunnelling through two localised states.** Figure 2a exhibits an additional feature in the measured $G(V_b, V_g)$ data. This arises from a more complex tunnelling process involving state A and a nearby localised state B with spatial coordinates $z_B$ and energy $E_B$. This process, in which a tunnelling electron makes three sequential steps, Gr → A → B → Gr, accounts for the broader peak in conductance highlighted by the loci of black dots in Fig. 2a, as explained at the end of this section. This additional contribution to the current flow is initiated when the bias and gate voltages are tuned so that states A and B are energetically aligned, $E_A = E_B$, allowing electrons to tunnel through the barrier in three steps, as shown schematically by the horizontal arrows in the top right inset of Fig. 2b. The two levels are aligned when $E_A^0 - E_B^0 = eF_b(z_A - z_B)$. Sequential tunnelling only occurs when the energies of the levels are aligned with each other and are located between $\mu_b$ and $\mu_t$. Therefore, the sequential tunnelling feature disappears when its locus intersects with that of the sharper conductance peak $E_A = \mu_b$ corresponding to the onset of tunnelling through state A alone (see horizontal dashed white arrow at the top of Fig. 2a, b).

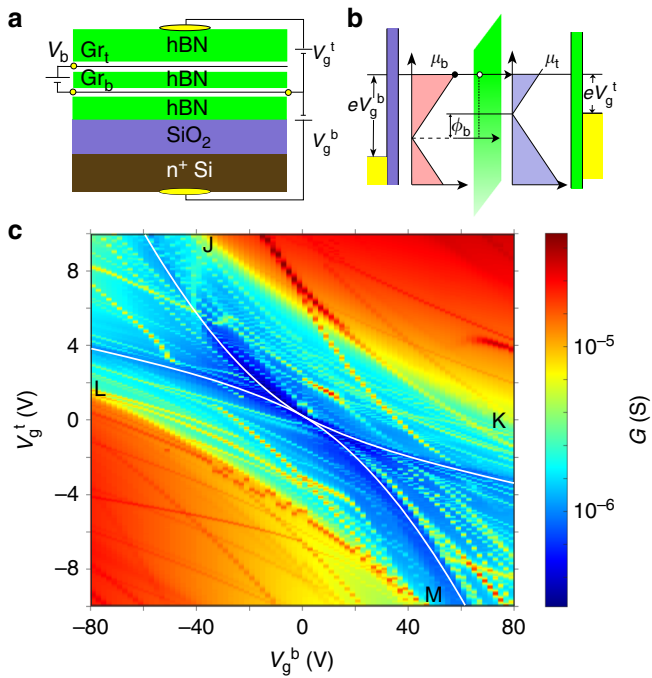

**Fig. 3** Colour map of differential conductance for Device 2 as a function of the two gate voltages. **a** Diagram of the double gated Device 2 showing the layer and voltage configuration where $V_b$, $V_g^t$ and $V_g^b$ are the applied bias, top gate and bottom gate voltages. **b** Schematic diagram showing the alignment of the chemical potentials in the bottom and top layers, $\mu_b$ and $\mu_t$, when $V_b = 0$. **c** Colour map (log-scale) of $G\left(V_g^b, V_g^t\right)$ when the bias voltage, $V_b = 0$, solid white curves show calculated positions when the chemical potential in the top and bottom electrodes intersect the Dirac points in those layers. J,K and L,M indicate the loci where there is a sharp increase in conductance corresponding to the threshold where electrons can tunnel directly between the bands of the two twisted graphene electrodes

The peak in conductance due to this three-step process corresponds to a step-like increase of current, which means that the current channel remains open when $E_B > E_A$. This requires an inelastic tunnelling process in which the electron loses energy as it tunnels between states A and B, see lower right inset in Fig. 2b. Such a process can occur by emission of a phonon[53] or else by an electron–electron interaction process analogous to Auger scattering whereby the tunnelling electron transfers the required excess energy to a free electron in one or other of the nearby graphene electrodes.

Comparison of Fig. 2a, b shows that the inclusion of this inelastic tunnelling process in our model (see Supplementary Note 3) provides an excellent fit to the data when we set the following parameters for state B: $E_B^0 = 0.02$ eV and $z_B = d$. These values imply that state B is situated near to the top graphene layer. Such a state could arise from an impurity or defect close to, or within, the top graphene layer, or from a local perturbation of the electronic states of this layer due to the close proximity of state A, giving rise to a peak in the local density of electron states of the top graphene electrode at an energy $E_B$[54]. Further evidence for this local enhancement is provided by the increased strength of the conductance peak associated with tunnelling through state A only at the intersection between the three and two-step processes, see the strong red contour highlighted by horizontal white arrow observed in the measured data, Fig. 2a, and confirmed in our model calculation, Fig. 2b. This observation of a three-step tunnelling transition process is of topical interest as it is an example of a percolation process, which has been recently reported in refs. [43,44].

Comparison of Fig. 2a, b also shows that the model successfully predicts the larger linewidth, $\Delta V_{AB} \sim 70$ mV, of the 3-step tunnelling peak compared to $\Delta V_A = 20$ mV for the peak arising from tunnelling through state A only. This increased broadening arises due to the addition of the linewidths of the two states. Note that the region of suppressed conductance (dark blue) predicted by the model, and the minimum of the double peak in the measured conductance (indicated by the locus of black dots in Fig. 2a) is fully consistent with the intersection of $E_A$ with the Dirac point in the top graphene layer, leading to a suppression in the number of electrons in the graphene layer available for sequential tunnelling.

**Position and energy spectroscopy of the localised states.** We now consider the current–voltage characteristics of a second type of device, Device 2, which has two-independent gate electrodes. The schematic diagram in Fig. 3a shows the layer and gate configuration. For this device, we observe a larger number (~50) of conductance peaks than for Device 1. The double gate arrangement provides further control over the electrostatics of the device. It allows us to select the particular combination of $\mu_b$ and $\mu_t$ required for electron tunnelling through a given localised state, see schematic diagram in Fig. 3b. The top gate is separated from the upper graphene layer by an insulating hBN barrier layer with thickness $d_g^t$, see schematic diagram. The doped Si substrate is used as the bottom gate electrode and is insulated from the lower graphene electrode by a SiO$_2$ surface layer and the thinner hBN bedding layer with a total thickness $d_g^b$ on which the lower graphene electrode, Gr$_b$, is mounted, see Fig. 3a. The lattices of the two monolayer graphene electrodes are misaligned by a small twist angle, $\theta$. The active area for electron tunnelling in this device is ~25 μm$^2$. The larger number of localised states observed in Device 2 may be due in part to the more complex processing required for this heterostructure.

Using a combination of conventional lock-in amplification and 4-probe DC measurements, we measured the tunnel current, $I$, and differential conductance with a small amplitude AC modulation voltage $\Delta V = 1$ mV at zero DC bias ($V_b = 0$) over a range of $V_g^b$ and $V_g^t$. This allows us to determine spectroscopically the energies of the localised states in the hBN tunnel barrier.

Figure 3c maps out the positions of the conductance peaks at zero applied bias voltage $G(V_b = 0)$ over a wide range of $V_g^b$ and $V_g^t$. When $V_b = 0$, the chemical potentials in the top and bottom graphene electrodes are aligned in energy, i.e. $\mu_b = \mu_t + \phi_b$. The electrostatic potential drop across the barrier, $\phi_b$, is strongly dependent on the two gate voltages. Figure 3c reveals a broad, dark blue cross-shaped region of very low conductance $G \lesssim 10^{-6}$ S. In this region, the Fermi energy in either the top or bottom graphene electrodes is close to the Dirac point in that layer (i.e. either $\mu_t$ or $\mu_b \approx 0$), where the density of states is low. The white loci show the calculated values of $V_g^b$ and $V_g^t$ when $\mu_t = 0$ and $\mu_b = 0$ using the electrostatic model presented in Supplementary Note 1, with $d_g^t = 21$ nm, $d_g^b = 310$ nm and $d = 1$ nm. The calculated loci show good agreement with the location of the measured conductance minima, thus confirming the accuracy of our model. Our model shows that at zero bias and zero gate voltages, the chemical potentials of the two graphene layers are within $40 \pm 10$ meV of their Dirac points corresponding to a hole doping level of $\sim 2 \times 10^{15}$ m$^{-2}$.

Figure 3c also reveals a sharp change in conductance from low to high (blue through yellow to orange) with well-defined loci, extending from J to K and from L to M in Fig. 3c. These correspond to the threshold at which electrons can tunnel directly between the two twisted graphene electrodes with conservation of momentum and energy[10] (i.e. not through localised states). The threshold condition is given by $\mu_b = \mu_t + \phi_b = (\phi_b \pm \Delta K v_F \hbar)/2$.

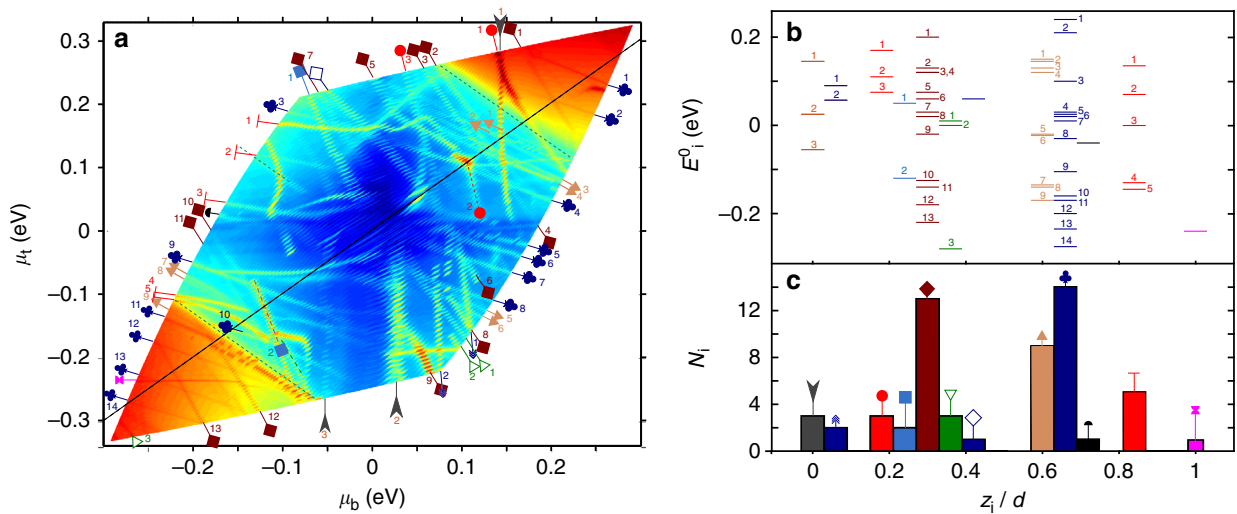

**Fig. 4** Energies and spatial locations of localised states within the hBN tunnel barrier as determined from the differential conductance measurements. **a** Colour map of the dependence of the zero bias differential conductivity as a function of the chemical potentials in the bottom and top graphene layers $G(\mu_b, \mu_t)$: symbols and associated lines highlight peaks corresponding to localised states with similar position, $z_i$, used to construct the histogram in **c**. The dashed line shows where $\mu_b = \mu_t$ and thus the electric field in the tunnel barrier, $F_b = 0$. **b** Horizontal lines show, on the y-axis, $E_i^0$, which are the energies of the localised states measured relative to the Dirac point of the bottom graphene layer in the flat band condition. The $z_i/d$ indicates their coordinate position within the hBN barrier. **c** Histogram showing the distribution of the positions of the localised states in the barrier. The widths of the lines in **b** and the bars in **c** reflect the uncertainty in the value of $d\mu_t/d\mu_b$

Using our model, we determine the misalignment of the in-plane wavevector between the Dirac points of the top and bottom layers: $\Delta K = 8\pi \sin(\theta/2)/3a$, where $a$ is the lattice constant of graphene. This provides a measure of $\theta = 2° \pm 0.5°$.

**Discussion**

We now consider the sharply defined curved loci of conductance peaks observed in the blue regions of the colour map in Fig. 3c where band-to-band tunnelling is suppressed, and also those in the yellow–red regions where the conductance peaks are superimposed on the high conductance regions that arise from momentum conserving, direct band-to-band, tunnel transitions. Each locus is due to resonant tunnelling through an energy level of a localised state and occurs when $\mu_b = \mu_t + \phi_b = E_i$, where $E_i$ is the energy of the state relative to the Dirac point of the bottom graphene electrode. To analyze the data in more detail, we remap $G(V_g^t, V_g^b)$ into a more useful colour plot of $G(\mu_t, \mu_b)$, using our electrostatic model, see Fig. 4a. As $E_i = E_i^0 + eF_b z_i$, we determine both $E_i^0$ and $z_i$ for each state. When $\phi_b = \mu_b - \mu_t = 0$ (shown by the black dashed line) the energy level of a given state, $i$, aligns with the chemical potential of the bottom graphene electrode so that $E_i^0 = \mu_b$, thus determining $E_i^0$ at the point when the peak trace crosses the black dashed curve. Whereas some of the loci of the conductance peaks in Fig. 4a are distinctly curved, most of them are approximately straight lines given by the relation

$$\mu_b(1 - z_i/d) = E_i^0 - \mu_t z_i/d. \quad (1)$$

Equation (1) and the colour map in Fig. 4a therefore allows us to determine the $z_i$ and $E_i^0$ values of each localised state from the gradient of the locus, $d\mu_b/d\mu_t$, and its position on the map. The results are shown in Fig. 4b, c. The width of each segment of the histogram in Fig. 4c indicates the accuracy, $\Delta z_i = 0.06$ nm, with which the position of each localised state is determined and reflects the uncertainty in the value of $d\mu_t/d\mu_b$. The histogram gives the number distribution, $N_i$ of states with respect to their position coordinate, $z_i$, within the hBN barrier.

Each bin of the histogram in Fig. 4c has a symbol and colour with which we label each conductance peak locus in Fig. 4a. Note that several peaks in $G$ have the same gradient and therefore the same value of $z_i/d$, within experimental error. For example, the peak at $z_i/d \approx 0.65$ with $N_i = 14$, includes a group of 4 conductance peaks, each labelled with a club-shape, numbered 11–14, in the lower left section of the plot and another group, clubs 5–8, in the lower right section. A second peak occurs when $z_i/d \approx 0.3$ (diamonds) corresponding to $N_i = 13$. The measurements therefore reveal that more than half of the detected states are located at or close to the two atomic layers that form the hBN barrier corresponding to $z_i/d \approx 0.3$ and $0.65$. We also find a number of states which appear to be located interstitially e.g. the five states at $z_i/d \sim 0.85$. Others with $z_i/d \approx 0$ and $\approx 1$ appear to be located close to the two monolayer graphene electrodes, possibly due to defects in or close to their lattices.

Figure 4b plots the energy, $E_i^0$, relative to the Dirac point of the bottom graphene layer at zero bias and gate voltages, and the binned position of each localised state in the barrier. To obtain the data shown in Figs. 3 and 4 we apply strong electric fields of up to a limit of $\sim\pm 300$ mV/nm across the barrier. This avoids the danger of electrical breakdown but limits our study to those localised states with energies, $E_i^0$, in the range $-0.3$ to $0.3$ eV. Previous studies indicate that the top of the valence band of hBN and the Dirac point of graphene are located at energies of $7.7 \pm 0.5$ eV[55] and $4.6 \pm 0.1$ eV[56,57] respectively, below the vacuum level. Based on these estimates, we determine that the group of localised states measured here are located in the mid-gap energy range between $2.8 \pm 0.5$ and $3.4 \pm 0.5$ eV above the valence band edge of the hBN barrier, with an average density of states of $\sim 3$ $\mu m^{-2}$ eV$^{-1}$.

Our measurements indicate that the areal density of tunnel-active defects in our devices is quite small $\lesssim 10^{12}$ m$^{-2}$, around 4 orders of magnitude smaller than the electron sheet densities in the graphene electrodes at zero bias and gate voltages. The average in-plane separation of the localised states is $\sim 1$ μm. These states are located at different depths within the thin hBN barrier layer and their energy levels appear to be distributed randomly over the energy range of $\sim 0.6$ eV that is accessible with these devices. However, the observation of the three-step electron tunnelling process requires that some localised states are in close proximity to each other, separated by ~1 nm.

In summary, we have observed resonant electron tunnelling between graphene monolayers through individual localised states in the hBN tunnel barrier. Our theoretical model determines the energy, linewidth, tunnel coupling coefficients and spatial coordinate of individual localised states in the barrier region. A three-step percolative inelastic process is also observed. These results may provide useful insights into the future exploitation and control of electron tunnelling through localised states in hBN.

## Methods

**Fabrication**. The devices were fabricated by a conventional dry-transfer procedure, the graphene and hBN layers were mechanically exfoliated onto the Si/SiO$_2$ substrate. Cr/Au contact pads were independently mounted on the single and bilayer graphene electrodes. Finally, the top hBN capping layer was covered by a Cr/Au layer of cross-sectional area 15 μm$^2$; this served as a top gate electrode. Further details of device fabrication can be found in ref. [10].

## Data availability

The datasets generated during this study are available from the corresponding authors on reasonable request.

### Acknowledgements
EU Graphene Flagship Program, European Research Council Synergy Grant Hetero2D, the Royal Society, Engineering and Physical Research Council (UK, grants EP/N007131/1 and EP/N010345/1), US Army Research Office (W911NF-16-1-0279). E.E.V. acknowledge support from Russian Science Foundation (17-12-01393), S.V.M. from NUST "MISiS" (K2-2017-009) and Yu.N.K from RAS Presidium Program N4 (task 007-00220-18-00). M.T.G acknowledges use of HPC Hydra at Loughborough University.

### Author contributions
The project was conceived by M.T.G., E.E.V., A.Mishchenko., A.K.G., K.S.N. and L.E. The measurements were carried out by E.E.V., D.G., A.Mishchenko., Z.W., Yu.N.K., S.V. M., O.M., A.P. and L.E. The theoretical model was devised by M.T.G., J.R.W., T.M.F., V.I. F. and L.E. The devices were fabricated by A.Misra, Y.C., M.H., and K.S.N. The hBN was prepared and supplied by K.W. and T.T.. The data were analysed by M.T.G., E.E.V., D.G., A.Mishchenko, K.S.N. and L.E. The manuscript was prepared by M.T.G., L.E. and K.S.N., with input from the other authors.

### Additional information
**Supplementary information** accompanies this paper at https://doi.org/10.1038/s42005-018-0097-1.

**Competing interests:** The authors declare no competing interests.

**Reprints and permission** information is available online at http://npg.nature.com/reprintsandpermissions/

**Publisher's note:** Springer Nature remains neutral with regard to jurisdictional claims in published maps and institutional affiliations.